\documentclass[12pt]{article}
\usepackage{a4}
\usepackage[cp1251]{inputenc}
\usepackage[english, ]{babel}
\usepackage[dvips]{graphicx}
\usepackage{graphicx}
\newcommand{\dis}{\displaystyle}

\begin{document}
\title{On generation of Abelian magnetic fields in $SU(3)$ gluodynamics at high temperature}
\author{V.V.Skalozub, A.V.Strelchenko\\
Dniepropetrovsk National University\\
St. Naukova 13, Dniepropetrovsk  49050, Ukraine\\
skalozub@ff.dsu.dp.ua, strel@ff.dsu.dp.ua}
\date{}
\maketitle
\thispagestyle{empty}
\begin{abstract}
The vacuum state of $SU(3)$ gluodynamics at high temperature is investigated.
A consistent approach including the calculation of either the spontaneously
generated constant chromomagnetic isotopic $H_{3}$ and hypercharge $H_{8}$
fields or the polarization operator of charged gluons in this background is
applied. It is shown within the effective potential including the one-loop
plus daisy diagrams  that the specific values of the fields deliver a global
minimum to free energy. The spectrum of the transversal charged modes is
stable at high temperature  due to a gluon magnetic mass accounting for the
vacuum  fields.  This leads to stable chromomagnetic fields in the
deconfinement phase of QCD. A comparison with the results of other
approaches is done.
\end{abstract}

\section{Introduction}

The deconfinement phase transition remains the most  topical problem of QCD
at finite temperature.  Nowadays a general believe is that the formation of
the magnetic monopole condensate at low temperature and its evaporation at
high temperature are responsible for this phenomenon (see, for instance,  the
papers \cite{Pol} - \cite{Cho} and references therein). This scenario  has been
investigated  either on a lattice \cite{Pol} - \cite{Bl} or in a continuum quantum field
theory \cite{Cho}. The present day status of this problem is characterized
by the fact that the results obtained by the former method are in agreement
with that of  the latter one and  complement each other, although  some
discrepancies exist. The most important discrepancy concerns  the properties
of the high temperature phase. As it is well  known from lattice
calculations, due to asymptotic freedom  at high temperature  the deconfining
phase is  to be a  gas of  free quarks and gluons - quark-gluon plasma.
No other  macroscopic parameters except temperature are expected.   However,
in the continuum calculations  in Refs.\cite{EO}, \cite{SVZ},
\cite{Skal}, \cite{SB}   the generation of the classical
chromomagnetic field of  order $ gH \sim g^4 T^2 $ was observed.
This spontaneously created  field is a reflection of  infrared dynamics
of the non-Abelian gauge fields at finite temperature.
In the papers \cite{EO}, \cite{Skal} the creation  of  the field
was considered. The field stabilization due to the electrostatic potential
(so-called the $A_0$ condensate) \cite{SVZ} and  radiation corrections
to the charged gluon spectrum \cite{SkSt1} has been investigated for the
$SU(2)$ gluodynamics. According to  the picture derived in these  papers the
vacuum at high temperature is to be a stable,  magnetized state. The noted
discrepancy  as well as some properties of the vacuum at high temperature
have been  discussed recently by Meisinger and Ogilvie  \cite{MO}.
To eliminate the classical  field in the deconfiming phase these authors have
introduced a gluon magnetic mass  on  heuristic grounds. Then it was observed
that to have a zero  field at high temperature the value of the magnetic mass
substituted into the one-loop effective potential (EP) must be of  order
$\sim g^2 T$. However,  the  spontaneous creation of  the chromomagnetic
field is mutually related  with  infrared properties of the non-Abelian gauge
fields \cite{Skal}, that  should be taken into consideration.
This  important  aspect of the gauge field dynamics at high temperature needs
in a more detailed investigation. Moreover, the correlation corrections
accounting for long distance effects should be included to find a consistent
picture at high temperature \cite{Kal}.

In the present paper the restored phase of QCD at high temperature is
investigated within an approach including  two stages  of calculation.
First,   the consistent  EP of   the Abelian constant chromomagnetic fields -
isotopic , $H_3$, and hypercharge, $H_8$ taking into consideration the
one-loop and the daisy diagrams, which include the gluon magnetic mass
insertions, is computed and the field configuration which  is spontaneously
generated is determined.  This potential is real due to the daisies of the
charged gluons, which cancel the  imaginary part  entering the one-loop part
of the EP.  As it is occurred,   a specific combination of   both fields is
formed. Second, the  one-loop polarization  operator  (PO) of  charged gluons
in these  external fields is calculated. It is averaged   over the gluon tree
level states in the fields in order to find the radiation corrections to the
spectrum. In this way  the effective Debye's and magnetic  masses of gluons
are derived. Then the vacuum magnetic field  strengths are used to
check whether or not the charged gluon spectrum (and therefore the magnetized
vacuum) at finite temperature is stable. As it is found, this is the case and
the non trivial vacuum is favorable at high temperature in the wide interval
of temperature above the decofinement transition temperature $T_d$.  Hence
we come to the conclusion that  the scenario with the magnetized vacuum
generated due to infrared dynamics of  gauge fields at high temperature
results in a  consistent  picture. The higher loop corrections can be
included  perturbatively within an external field problem.

The paper is organized as follows.
In Sect.2 the charged sector of the
$SU_{c}(3)$ gluodynamics is introduced. In Sect.3 the
generation of the external chromomagnetic fields is considered
and the field strengths are derived.
 In Sect.4 we calculate the PO in the external Abelian
chromomagnetic fields at finite temperature and carry out
the high temperature expansion of the one-loop radiation corrections
 to the Landau levels. The last section is devoted to
 the discussion of results obtained and possible applications.

\section{The model}

The Lagrangian of $SU(3)$ gluodynamics reads \cite{Yn}
 \begin{equation} \label{one}
 L=-\frac{1}{4}F_{\mu\nu}^{a}F_{\mu\nu}^{a}+L_{gf}+L_{gh},
 \end{equation}
where $ F_{\mu\nu}^{a}=\partial_{\mu} A_{\nu}^{a}-\partial_{\nu}
A_{\mu}^{a}-g f^{abc}A_{\mu}^{a}A_{\nu}^{a}$ is the field strength
tensor, $A_{\mu}^{a}$ is a potential of the gluon field, $f^{abc}$
are the $SU_{c}(3)$ structure constants, $a=1,...,8$. The metric is
chosen to be Euclidian to consider the theory at $T\neq 0$ in
the imaginary time formalism.
The external chromomagnetic field is introduced by dividing the gluon
field $A_{\mu}^{a}$ into the sum of the classic background field
$B_{\mu}^{a}$ and the quantum field $Q_{\mu}^{a}$ ,
\begin{equation}\label{two}
A_{\mu}^{a}=B_{\mu}^{a}+Q_{\mu}^{a}.
\end{equation}
We choose the external potential in the form
$B_{\mu}^{a}=\delta^{a3}B_{3\mu}+\delta^{a8}B_{8\mu}$,
where $B_{3\mu}=H_{3}\delta_{\mu2}x_{1}$
and
$B_{8\mu}=H_{8}\delta_{\mu2}x_{1}$
correspond to constant chromomagnetic fields directed along the
third axis of the Euclidean space and  $a=3$ and $a=8$ of the colour
$SU_{c}(3)$-space, respectively: $F_{\mu\nu}^{a ext}=\delta^{a3}F_{3\mu\nu}+\delta^{a8}F_{8\mu\nu}$,
$F_{c12}=-F_{c21}=H_{c}$, $c=3,8$. The gauge fixing term in the Eq.(\ref{one}) is
\begin{equation}\label{three}
L_{gf}=-\frac{1}{2}(\partial_{\mu}Q_{\mu}^{a}
       +g f^{abc}B_{\mu}^{b}Q_{\mu}^{c})^{2}
\end{equation}
and $L_{gh}$ represents the ghost Lagrangian.
The components $Q_{\mu}^{a}$ with $a=1,2,4,5,6,7$ correspond to the charged
gluons. It is convenient to introduce the "charged basis" of fields
 $Q_{\mu}^{a}$ ($a=1,2,4,5,6,7$) by the expressions
\begin{equation}\label{four}
W_{1\mu}^{\pm}=\frac{1}{\sqrt{2}}(Q_{\mu}^{1}\pm iQ_{\mu}^{2}),
~ W_{2\mu}^{\pm}=\frac{1}{\sqrt{2}}(Q_{\mu}^{4}\pm iQ_{\mu}^{5}),
~ W_{3\mu}^{\pm}=\frac{1}{\sqrt{2}}(Q_{\mu}^{6}\pm iQ_{\mu}^{7}).
\end{equation}
After simple algebra, one obtains the Lagrangian of the charged gluons in
the form
\begin{eqnarray}\label{five}
&L_{ch.gl.}=\sum_{r=1}^3\left(-\frac{1}{2} W_{r\mu\nu}^{+}W_{r\mu\nu}^{-}
-(D^{\ast}_{\mu}W_{r\mu}^{+})(D_{\nu}W_{r\nu}^{-})
-\frac{1}{2}c_{r}g^{2}W_{r\mu}^{+}W_{r\nu}^{-}W_{r\lambda}^{+}W_{r\rho}^{-}
\Gamma_{\mu\nu\lambda\rho}\right)&\nonumber\\
&~~~~~~~~~~~~~+i  g(F_{3\mu\nu}+Q_{\mu\nu}^{3}) W_{1\mu}^{+}W_{1\nu}^{-}
+igQ_{\mu}^{3}(W_{1\nu}^{+}(\partial_{\mu}W_{1\nu}^{-}-\partial_{\nu}W_{1\mu}^{-})
-(h.c.)) &\nonumber\\
&+i \sqrt{\frac{3}{2}} g(\lambda_{+}F_{8\mu\nu}
+Q_{\mu\nu}^{8}+\frac{1}{\sqrt{6}}Q_{\mu\nu}^{3}) W_{2\mu}^{+}W_{2\nu}^{-}~~~~~~~~~~~~~~~~~~ &\nonumber\\
&+i\sqrt{\frac{3}{2}}g(Q_{\mu}^{8}+\frac{1}{\sqrt{6}}Q_{\mu}^{3})
(W_{2\nu}^{+}(\partial_{\mu}W_{2\nu}^{-}-\partial_{\nu}W_{2\mu}^{-})
-(h.c.))
~~~&\nonumber\\
&+i \sqrt{\frac{3}{2}} g(\lambda_{-}F_{8\mu\nu}
+Q_{\mu\nu}^{8}-\frac{1}{\sqrt{6}}Q_{\mu\nu}^{3}) W_{3\mu}^{+}W_{3\nu}^{-}~~~~~~~~~~~~~~~~~~ &\nonumber\\
&~~~~~~+i\sqrt{\frac{3}{2}}g(Q_{\mu}^{8}-\frac{1}{\sqrt{6}}Q_{\mu}^{3})
(W_{3\nu}^{+}(\partial_{\mu}W_{3\nu}^{-}-\partial_{\nu}W_{3\mu}^{-})
-(h.c.))
+L_{gh},&
\end{eqnarray}
where
$W_{r\mu\nu}^{+}=D_{r\mu}^{\ast}W_{r\nu}^{+}-D_{r\nu}^{\ast}W_{r\mu}^{+}$,
$W_{r\mu\nu}^{-}=D_{r\mu}W_{r\nu}^{-}-D_{r\nu}W_{r\mu}^{-}$,
$D_{r=1~\mu}=\partial_{\mu}+i gB_{3\mu}$,
$D_{r=2,3~\mu}=\partial_{\mu}+i\sqrt{\frac{3}{2}}\lambda_{\pm} gB_{8\mu}$,
are covariant
derivatives,
$\Gamma_{\mu\nu\lambda\rho}=\delta_{\mu\nu}\delta_{\lambda\rho}
-\delta_{\mu\lambda}\delta_{\nu\rho} $,
 $$\lambda_{\pm}=1\pm\frac{1}{\sqrt{6}}\frac{H_{3}}{H_{8}},$$
$c_{r}=1,\frac{21}{12},\frac{15}{12}$
for $r=1,2,3$, respectively.
The Lagrangian (\ref{five}) is a starting point of our analysis.

\section{The spontaneous generation of chromomagnetic fields }
First, let us investigate the spontaneous vacuum
magnetization in the high temperature $SU_{c}(3)$
gluodynamics, charged sector of which is  described by the Lagrangian (\ref{five}).
For this purpose we apply the method of the effective Lagrangian.

The effective Lagrangian of constant chromomagnetic
fields $H_{3}$ and $H_{8}$
at finite temperature
can be written in the form:
\begin{equation}\label{six}
L_{eff}=L^{(1)}+L^{(ring)}+\ldots,~~~~~~~~~~~~~~~~~~~~
\end{equation}
where the first term represents the one-loop contribution of the
charged gluons:
\begin{eqnarray}\label{seven}
&\dis L^{(1)}=-\frac{gH_{3}}{2\pi\beta}\sum_{l=-\infty}^\infty~ \int\limits_{-\infty}^\infty \frac{d p_{3}}{2 \pi}
\sum_{n,\sigma}\ln[\beta^{2}G_{r=1}^{-1}(p_{3},H_{3},T)]
 ~~~~~~~~~~~~~~&\nonumber\\
 &\dis-~\sqrt{\frac{3}{2}}
\lambda_{+}\frac{gH_{8}}{2\pi\beta}\sum_{l=-\infty}^\infty~ \int\limits_{-\infty}^\infty \frac{d p_{3}}{2 \pi}
\sum_{n,\sigma}\ln[\beta^{2}G_{r=2}^{-1}(p_{3},H_{3},H_{8},T)]
&\nonumber\\
 &\dis-~\sqrt{\frac{3}{2}}
\lambda_{-}\frac{gH_{8}}{2\pi\beta}\sum_{l=-\infty}^\infty~ \int\limits_{-\infty}^\infty \frac{d p_{3}}{2 \pi}
\sum_{n,\sigma}\ln[\beta^{2}G_{r=3}^{-1}(p_{3},H_{3},H_{8},T)]
.&
\end{eqnarray}
Here $r$ marks the index of the charged basis (\ref{four}), $G_{r}$
is the corresponding propagator of the charged gluons in the
external fields $H_{3}$ and $H_{8}$. The second term in Eq.
(\ref{six})
presents the contribution of daisy or ring diagrams of the charged
gluons,
\begin{eqnarray}\label{eight}
&\dis L^{(ring)}_{ch}=-\frac{gH_{3}}{2\pi\beta}\sum_{l=-\infty}^\infty~
\int\limits_{-\infty}^\infty \frac{d p_{3}}{2 \pi}
\sum_{n,\sigma}\ln[1+
G_{r=1}(\epsilon^{2}_{n},H_{3},T)\Pi^{r=1}(H_{3},T,n,\sigma)]
&\nonumber\\
 &\dis-~\sqrt{\frac{3}{2}}
\lambda_{+}\frac{gH_{8}}{2\pi\beta}\sum_{l=-\infty}^\infty~ \int\limits_{-\infty}^\infty \frac{d p_{3}}{2 \pi}
\sum_{n,\sigma}\ln[1+&\nonumber\\
&~~~~~~~~~~~~~~~~~~~~~~~~~~~~~~~~~~~~~~~~
+~G_{r=2}(\epsilon^{2}_{n},H_{3},H_{8},T)\Pi^{r=2}(H_{3},H_{8},T,n,\sigma)]
&\nonumber\\
 &\dis-~\sqrt{\frac{3}{2}}
\lambda_{-}\frac{gH_{8}}{2\pi\beta}\sum_{l=-\infty}^\infty~ \int\limits_{-\infty}^\infty \frac{d p_{3}}{2 \pi}
\sum_{n,\sigma}\ln[1+&\nonumber\\
&~~~~~~~~~~~~~~~~~~~~~~~~~~~~~~~~~~~~~~~~
+~G_{r=3}(\epsilon^{2}_{n},H_{3},H_{8},T)\Pi^{r=3}(H_{3},H_{8},T,n,\sigma)]
,&
\end{eqnarray}
and of neutral gluons,
\begin{eqnarray}\label{nine}
&\dis L^{(ring)}_{neut}=-\frac{1}{2\beta}\sum_{l=-\infty}^\infty~
\int\limits_{-\infty}^\infty \frac{d p_{3}}{2 \pi}
\ln[\omega_{l}^{2}+\overline{p}^{2}+
\Pi'(H_{3},H_{8},T)]
&\nonumber\\
&\dis-\frac{1}{2\beta}\sum_{l=-\infty}^\infty~
\int\limits_{-\infty}^\infty \frac{d p_{3}}{2 \pi}
\ln[\omega_{l}^{2}+\overline{p}^{2}+
\Pi''(H_{3},H_{8},T)]
.&
\end{eqnarray}
These expressions include the PO of charged gluons
averaged over physical states, which are also
dependent on $H=H_{3},H_{8}$, the level number $n=0,1,\ldots$,
the spin projection $\sigma=\pm1$, and the Debye
masses of the  neutral gluons
$Q^{3}_{\mu}$,~$Q^{8}_{\mu}$,
$$\Pi'(H,T)=\Pi'_{00}(k=0,H,T),$$
$$\Pi''(H,T)=\Pi''_{00}(k=0,H,T),$$
respectively.
The averaged values of charged gluon PO taken in
the state $n=0$ and $\sigma=+1$ give the magnetic masses of the
transversal modes.
The quantities $\Pi'(H,T)$
and $\Pi''(H,T)$ are the
zero-zero components of the corresponding neutral gluon
polarization operators calculated in the external fields
$H=H_{3},H_{8}$ at finite temperature and taken at zero momentum.
The ring contribution to the $L_{eff}$ has to be calculated when the vacuum
magnetization at non-zero temperature is investigated.
These diagrams account for long range correlations at finite
temperature \cite{SB}.

The detailed evaluations of the one-loop effective Lagrangian in
finite-temperature
$SU(2)$ gluonodynamics
have been carried
out in Refs.\cite{Skal},\cite{SB}.
Performing them for our case we
arrive at the following result for the high-temperature limit of
$L^{(1)}$:
\begin{eqnarray}\label{ten}
& L^{(1)}=\dis-\frac{H_{3}^{2}}{2}
-\frac{11}{32}\frac{g^{2}}{\pi^{2}}H_{3}^{2}\ln[\frac{T}{\mu}]
+\frac{1}{3\pi}(gH_{3})^{\frac{3}{2}}\frac{T}{3\pi}&\nonumber\\
&
\dis
-\frac{H_{8}^{2}}{2}-\frac{11}{16}\frac{g^{2}}{\pi^{2}}H_{8}^{2}\ln[\frac{T}{\mu}]
+(
\lambda_{+}^{\frac{3}{2}}
+|\lambda_{-}|^{\frac{3}{2}})
(\frac{3}{2})^{\frac{3}{4}}(gH_{8})^{\frac{3}{2}}\frac{T}{3\pi}&\nonumber\\
&
\dis+i[(gH_{3})^{\frac{3}{2}}
+(
\lambda_{+}^{\frac{3}{2}}
+|\lambda_{-}|^{\frac{3}{2}})
(\frac{3}{2})^{\frac{3}{4}}(gH_{8})^{\frac{3}{2}}]
\frac{T}{2\pi},&
\end{eqnarray}
where $T\gg \sqrt{gH_{3,~8}}\gg \mu$, $\mu$ is a renormalization
point.
 The imaginary part in the expression (\ref{ten}) signals
for the vacuum instability and must be considered carefully. Namely, as
it will be shown below, the inclusion of
ring diagrams , $L^{ring}$, leads to canceling the imaginary
parts so that the whole expression $L_{eff}$ becomes real. To see
this, let us consider the contribution of the ring
diagrams which correspond to the unstable modes of the charged gluons:

\begin{eqnarray}\label{eleven}
&\dis L^{(ring)}_{unstable}=-\frac{gH_{3}}{2\pi\beta}\sum_{l=-\infty}^\infty~
\int\limits_{-\infty}^\infty \frac{d p_{3}}{2 \pi}
\ln[1+(\omega_{l}^{2}+p_{3}^{2}-gH_{3})^{-1}\Pi^{r=1}(H_{3},T)]
~~~~~~~~~&\nonumber\\
&
\dis-~\sqrt{\frac{3}{2}}
\lambda_{+}\frac{gH_{8}}{2\pi\beta}\sum_{l=-\infty}^\infty~
 \int\limits_{-\infty}^\infty \frac{d p_{3}}{2 \pi}
\ln[1+(\omega_{l}^{2}+p_{3}^{2}-\sqrt{\frac{3}{2}}
\lambda_{+} gH_{8})^{-1}\Pi^{r=2}(H_{3},H_{8},T)]
&\nonumber\\
&
\dis-~\sqrt{\frac{3}{2}}
\lambda_{-} \frac{gH_{8}}{2\pi\beta}\sum_{l=-\infty}^\infty~
 \int\limits_{-\infty}^\infty \frac{d p_{3}}{2 \pi}
\ln[1+(\omega_{l}^{2}+p_{3}^{2}-\sqrt{\frac{3}{2}}
\lambda_{-} gH_{8})^{-1}\Pi^{r=3}(H_{3},H_{8},T)],
&
\end{eqnarray}
where $\omega_{l}=2\pi l T$,~~$l=0,\pm1,\ldots,$ are the Matsubara frequencies.
To obtain (\ref{eleven}) one has merely to put $n=0$ and
$\sigma=+1$ in the expression for $L^{ring}_{ch}$.
An elementary integration gives
\begin{eqnarray}\label{twelve}
&\dis L^{(ring)}_{unstable}=-\frac{gH_{3}T}{2\pi}
[\Pi^{r=1}(H_{3},T)-gH_{3}]^{\frac{1}{2}}&\nonumber\\
&
\dis-~\sqrt{\frac{3}{2}}
\lambda_{+} \frac{gH_{8}T}{2\pi}
[\Pi^{r=2}(H_{3},H_{8},T)-\sqrt{\frac{3}{2}}
\lambda_{+} gH_{8}]^{\frac{1}{2}}
&\nonumber\\
&
\dis-~\sqrt{\frac{3}{2}}
\lambda_{-} \frac{gH_{8}T}{2\pi}
[\Pi^{r=3}(H_{3},H_{8},T)-\sqrt{\frac{3}{2}}
\lambda_{-} gH_{8}]^{\frac{1}{2}}
&\nonumber\\
&
\dis-i[(gH_{3})^{\frac{3}{2}}
+(
\lambda_{+}^{\frac{3}{2}}
+|\lambda_{-}|^{\frac{3}{2}})
(\frac{3}{2})^{\frac{3}{4}}(gH_{8})^{\frac{3}{2}}]
\frac{T}{2\pi}.&
\end{eqnarray}
From Eq.(\ref{ten}) and (\ref{twelve}) it is seen that imaginary parts are
cancelled out in the total. The final effective Lagrangian $L_{eff}$
is real if the relations
$$\Pi^{r=1}(H_{3},T)>gH_{3},$$
$$\Pi^{r=2}(H_{3},H_{8},T)>\sqrt{\frac{3}{2}}
\lambda_{+}gH_{8}$$
and
$$\Pi^{r=3}(H_{3},H_{8},T)>\sqrt{\frac{3}{2}}
\lambda_{-}gH_{8}$$
hold.

In one-loop order the neutral gluon contribution is a trivial $H$-
independent constant which can be omitted. However, these fields
are long range states and give $H$-dependent effective
Lagrangian through the correlation corrections (\ref{nine}) depending on
the temperature and external fields. Below, only the longitudinal
neutral modes are included because their Debye's masses are
nonzero. The corresponding effective Lagrangian is easily
calculated and has the form \cite{SB}
\begin{eqnarray}\label{thirteen}
&\dis L^{(ring)}_{neut}=-\frac{T^{2}}{24}
[\Pi'(H_{3},H_{8},T)+\Pi''(H_{3},H_{8},T)]&\nonumber\\
&\dis+\frac{T}{12\pi}[\Pi'(H_{3},H_{8},T)+\Pi''(H_{3},H_{8},T)]^{\frac{3}{2}}&\nonumber\\
&\dis-\frac{1}{32\pi^{2}}[\Pi'(H_{3},H_{8},T)+\Pi''(H_{3},H_{8},T)]^{2}&\nonumber\\
&\dis(\log[4\pi T
[\Pi'(H_{3},H_{8},T)+\Pi''(H_{3},H_{8},T)]^{-\frac{1}{2}}+\frac{3}{4}-\gamma]).&
\end{eqnarray}
Evaluating the Debye masses of the neutral gluons $Q^{3}_{\mu}$,~$Q^{8}_{\mu}$
gives the following results (see for details \cite{SB}):
\begin{eqnarray}\label{fourteen}
&\dis
(m'_{D})^{2}=\Pi'_{00}(k=0,H,T)=\frac{2}{3}g^{2}T^{2}-\frac{g^{2}T}{\pi}\times
~~~~~~~~~~~~~~~~~~~~~~~~~~~~~~~~&\nonumber\\
&\dis\times[(gH_{3})^{\frac{1}{2}}
+(
\lambda_{+}^{\frac{1}{2}}
+|\lambda_{-}|^{\frac{1}{2}})
(\frac{3}{2})^{\frac{1}{4}}(gH_{8})^{\frac{1}{2}}]]&\nonumber\\
&\dis
(m''_{D})^{2}=\Pi''_{00}(k=0,H,T)=\frac{2}{3}g^{2}T^{2}-\frac{g^{2}T}{\pi}
(\lambda_{+}^{\frac{1}{2}}
+|\lambda_{-}|^{\frac{1}{2}})
(\frac{3}{2})^{\frac{1}{4}}(gH_{8})^{\frac{1}{2}}.&
\end{eqnarray}
Substituting expressions (\ref{fourteen})
into $L^{(ring)}_{neut}$ we obtain the correlation corrections due
to neutral gluons:
\begin{equation}\label{fifteen}
\dis L^{(ring)}_{neut}=\frac{g^{2}T^{3}}{24\pi}
[(gH_{3})^{\frac{1}{2}}
+2(
\lambda_{+}^{\frac{1}{2}}
+|\lambda_{-}|^{\frac{1}{2}})
(\frac{3}{2})^{\frac{1}{4}}(gH_{8})^{\frac{1}{2}}],
\end{equation}
where the $H$-independent terms were skipped. Thus, the vacuum
magnetization at high temperature $T\gg \sqrt{gH_{3,8}}$ will be
investigated within the following effective Lagrangian:
\begin{eqnarray}\label{sixteen}
& L^{(eff)}=\dis-\frac{H_{3}}{2}-\frac{H_{8}}{2}
-\frac{11}{32}\frac{g^{2}}{\pi^{2}}H_{3}^{2}\ln[\frac{T}{\mu}]
-\frac{11}{16}\frac{g^{2}}{\pi^{2}}H_{8}^{2}\ln[\frac{T}{\mu}]
~~~~~~~~~~~~~&\nonumber\\
&
\dis+[(gH_{3})^{\frac{3}{2}}
+(
\lambda_{+}^{\frac{3}{2}}
+|\lambda_{-}|^{\frac{3}{2}})
(\frac{3}{2})^{\frac{3}{4}}(gH_{8})^{\frac{3}{2}}]\frac{T}{3\pi}~~&\nonumber\\
&\dis-\frac{gH_{3}T}{2\pi}
[\Pi^{r=1}(H_{3},T)-gH_{3}]^{\frac{1}{2}}&\nonumber\\
&
\dis-~\sqrt{\frac{3}{2}}
\lambda_{+} \frac{gH_{8}T}{2\pi}
[\Pi^{r=2}(H_{3},H_{8},T)-\sqrt{\frac{3}{2}}
\lambda_{+} gH_{8}]^{\frac{1}{2}}
&\nonumber\\
&
\dis-~\sqrt{\frac{3}{2}}
\lambda_{-} \frac{gH_{8}T}{2\pi}
[\Pi^{r=3}(H_{3},H_{8},T)-\sqrt{\frac{3}{2}}
\lambda_{-} gH_{8}]^{\frac{1}{2}}
&\nonumber\\
&\dis+\frac{g^{2}T^{3}}{24\pi}
[(gH_{3})^{\frac{1}{2}}
+2(
\lambda_{+}^{\frac{1}{2}}
+|\lambda_{-}|^{\frac{1}{2}})
(\frac{3}{2})^{\frac{1}{4}}(gH_{8})^{\frac{1}{2}}]
+O(g^{3}).&
\end{eqnarray}
This expression includes the contributions of $L^{(1)}$ as well as
$L^{ring}_{unstable}$
and $L^{ring}_{neut}$.
Notice that the quantity $L^{ring}_{unstable}$ has
the order $g^{\frac{9}{4}}$
in coupling constant $g$ , whereas the order of $L^{ring}_{neut}$
is $g^{\frac{5}{2}}$.
In other wards, the contribution of the neutral gluons
does not play essential role in
generation of external fields and
this part
cannot be taken into account.
This is natural on general grounds because the neutral gluon field
is stable at the tree and the one-loop levels. So, one has not to
expect any role of this sector in the field generation.
Thus, in this
approximation, $L^{(eff)}$ is equal to the
effective Lagrangian (\ref{sixteen}) without the terms
in the last line.

 Our problem is divided into two separate parts: first, one has to
calculate the spontaneously generated fields in the vacuum and,
second, to compute $\Pi^{r}(H,T,n,\sigma)$, which are the average
values of the charged gluon PO taken in the tree level states.

To derive the strengths of the generated fields one has to solve
the set of the stationary equations:
$$\frac{\partial{L^{(eff)}}}{\partial{H_{3}}}=0,$$
$$\frac{\partial{L^{(eff)}}}{\partial{H_{8}}}=0.$$
There are three nontrivial solutions:
\begin{equation}\label{seventeen}
H_{3}=0,~~
H_{8}=(\frac{3}{2})^{\frac{3}{2}}~\frac{g^{3}T^{2}}{\pi^{2}}~~~~~~~~~~~~~~~~~~~~~~~~~
\end{equation}
\begin{equation}\label{eighteen}
H_{3}=\frac{1}{4}~(1+\frac{1}{\sqrt{2}})^{2}~\frac{g^{3}T^{2}}{\pi^{2}},~~
H_{8}=0~~~~~~~~~~~~~~~
\end{equation}
and
\begin{equation}\label{ninteen}
H_{3}=0.2976~\frac{g^{3}T^{2}}{\pi^{2}},~~
H_{8}=0.9989~(\frac{3}{2})^{\frac{3}{2}}~\frac{g^{3}T^{2}}{\pi^{2}}.
\end{equation}
The terms of $L^{eff}$ that depend on the magnetic masses of the transversal
modes are not included in the expressions
(\ref{seventeen})-(\ref{ninteen}) because their contributions are of
higher order in $g$.
Note that the latter of these configurations corresponds to the
minimum of the EP and we,therefore, conclude that
both chromomagnetic fields have to be arised spontaneously at high
temperature.

 In determining the solutions (\ref{seventeen}) -(\ref{ninteen}), the logarithmic terms
 $\sim \ln[\frac{T}{\mu}]$ were omitted as negligibly small.
 This approximation is appropriate for the region
 $T \geq T_{d}$, where $T_{d}$ is the deconfinement
 phase transition temperature. However,
 in the limit $T\rightarrow \infty$, the
 logarithmic terms become large and should be accounted for.
 To analyze the asymptotic region the solutions of the above
 stationary equations must be rewritten in terms of the effective
 coupling constant
 $$g_{eff}^{2}\approx (\frac{11}{16
 \pi^{2}}\ln[\frac{T}{\mu}])^{-1}.$$
For this purpose one has merely to eliminate the tree-level
 terms in the $L^{(eff)}$.
It turns out that, at
$T\gg T_{d}$, the field configuration
\begin{equation}\label{twenty}
gH_{3}=\frac{1}{4}~(1+\frac{1}{\sqrt{2}})^{2}~\frac{g_{eff}^{4}T^{2}}{\pi^{2}},~~
gH_{8}=0~~~~~~~~~~~~~~~
\end{equation}
delivers the global minimum of the EP.
That is only the isotopic chromomagnetic field is generated
at asymptotically high temperatures. If the temperature
decreases, the hypercharge $H_{8}$ field appears below some temperature $T_{0}$,
and in the deconfinement region $T_{0}>T \geq T_{d}$,
where the terms $\sim  \ln[\frac{T}{\mu}]$
are
small in comparison with the tree-level ones,
both fields are present.

\section{Gluon polarization operator}
The next question that must be answered is whether the
obtained in (\ref{ninteen}) chromomagnetic fields $H_{3}$ and
$H_{8}$ are stable. This sufficiently complicate
problem requires an explicit calculation of the polarization operator of the
charged gluons in the external fields $H_{3}$ and
$H_{8}$ .

In the one-loop approximation the PO of the
charged gluons is determined by the standard set of diagrams
in Fig. 1, where double wavy lines represent the Green function
$G_{r~\mu\nu} (x,y)$ for the charged gluons, dashed double lines
represent the Green function $D(x,y)$ for the charged ghost
fields. Thin wavy and thin dashed lines correspond to the neutral
gluon fields $Q_{\mu}^{3,8}$ and the neutral ghost fields $C_{3,8}$,
respectively. In the operator form the above Green functions are
 given by the expressions (in Feynmann's gauge)

\[G_{r=1~\mu\nu}(P)=-[P^2+2igF_{3\mu\nu}]^{-1},\]
\[G_{r=2,3~\mu\nu}(P)=-[P^2+\sqrt{6}i\lambda_{\pm} gF_{8\mu\nu}]^{-1},\qquad
D(P)=-\frac{1}{P^2}.\]

To calculate the PO we make use of the proper time
representation and the Schwinger operator formalism \cite{Sch}.
The PO of the charged gluons in a chromomagnetic field at nonzero
temperature can be written as
\begin{equation}\label{twentyone}
\Pi_{\mu\nu}^{r}=\frac{g^2}{\beta}~c_{r}\sum_{k_4}
\int\frac{d^3k}{(2\pi)^3} \Pi^{r}_{\mu\nu}(k,P),~~ r=1,2,3,
\end{equation}
where
\begin{eqnarray*}
\Pi^{r}_{\mu\nu}(k,P)&=&k^{-2}\Bigl\{\Gamma_{\mu\alpha,\rho}
G_{r~\alpha\beta}(P-k)\Gamma_{\nu\beta,\rho}+(P-k)_\mu
D(P-k)k_\nu+ \\
& &+k_\mu D(P-k)(P-k)_\nu+ \\
& &+k^2 \Bigl[G_{r~\mu\nu}(P-k)-2G_{r~\nu\mu}(P-k)+
\delta_{\mu\nu}G_{r~\rho\rho}(P-k)\Bigr]\Bigr\},\\
\Gamma_{\mu\alpha,\rho}&=&\delta_{\mu\alpha}(2P-k)_\rho+\delta_{\alpha\rho}
(2k-P)_\mu+\delta_{\mu\rho}(P+k)_\alpha,
\end{eqnarray*}
$\beta =\frac{1}{T}$, $k_{4} =\frac{2 \pi l}{\beta}$, $l=0, \pm1,
\pm2, \ldots$,
$P_{\mu}= i \partial_{\mu} +g B_{3\mu}$ for $r=1$
and $P_{\mu}= i \partial_{\mu} +
\sqrt{\frac{3}{2}}\lambda_{\pm} gB_{8\mu}$ for $r=2,3$,
respectively, the constant $c_{r}$ is defined above.
 We restrict our consideration to the case of a high
temperature limit. In Eq.(\ref{twentyone}) this limit corresponds to the
$l=0$ term in the sum over $k_{4}$ \cite{Kal}. To
evaluate the expression for the PO we used the Schwinger
proper-time method modified for the case of the high temperature (see Ref.
\cite{SkSt1} for details).
Thus, the average over the physical states values for the gluon PO and the
 Debye mass squared of charged gluons can be presented as

\begin{eqnarray}\label{twentytwo}
&\langle n,\sigma \mid \Pi^{r}_{ij} \mid
n,\sigma\rangle =\Pi^{r}(P_4=0,h_{r},T,n,\sigma)=&\nonumber\\
&\dis=\frac{g^{2}}{8 \pi^{3/2}\beta}~c_{r}
\int\limits_0^1 \frac{d u}{\sqrt{u}}
\int\limits_0^\infty \frac{d x}{\sqrt{x}}
\left[g_{r}h_{r}\Delta\right]^{-1/2}\times~~&\nonumber\\
&~~~~~~\times\exp\Bigl\{-(2n+1)[\rho-x(1-u)]-2y(1-u)\Bigr\}\Pi^{r}(x,u),&
\end{eqnarray}
\begin{eqnarray}\label{twentythree}
&\Pi_{44}^{r}(P_4=0,h_{r},T,n)=\dis
\frac{g^{2}}{8 \pi^{3/2}\beta}~c_{r}
\int\limits_0^1 \frac{d u}{\sqrt{u}}
\int\limits_0^\infty \frac{d x}{\sqrt{x}}
[g_{r}h_{r}\Delta]^{-1/2}~~&\nonumber\\
&~~~~~~~~~~~~~~~~~~~~~~~\times\exp\Bigl\{-(2n+1)[\rho-x(1-u)]\Bigr\}\widetilde{\Pi}^{r}(x,u),&
\end{eqnarray}
where $x=g_{r}h_{r}us$, $y=x\sigma$, $\sigma=\pm1$, $h_{r=1}=H_{3}$,
$h_{r=2}=\lambda_{+}H_{8}$,
$h_{r=3}=\lambda_{-}H_{8}$,
$g_{r=1}=g$, $g_{r=2,3}=\sqrt{\frac{3}{2}}g$,
$$\tanh \rho=\frac{(1-u)\sinh x}{(1-u)\cosh x+u\frac{\sinh x}{x}},$$
$$\Delta=(1-u)^2+2u(1-u)\frac{\sinh 2x}{2x}+u^2\frac{\sinh ^2x}{x^2}.$$
The explicit expressions of the functions $\Pi^{r}(x,u)$ and
$\widetilde{\Pi}^{r}(x,u)$ are complicate and have in general
the same forms
as in the case of $SU(2)$ theory considered in \cite{SkSt1}.
However, for our analysis we need only
the asymptotic expressions for $\Pi^{r}(x,u)$ and
$\widetilde{\Pi}^{r}(x,u)$ when $u \sim 1$,
and $x \gg 1$ that corresponds to the high-temperature expansion,
$\frac{gH}{T^{2}}\ll 1$ (see  \cite{SkSt1}).
Without loss of generality, the calculations can be carried out
in the reference frame $P_{3}=0$. Performing integrations we obtain
\begin{eqnarray}\label{twentyfour}
&\Pi^{r=1}(P_{4}=0, P_{3}=0, H_{3}, T, n,\sigma=+1)=
~~~~~~~~~~~~~~~~~~~~~~~~~~~~~~~~~~~~~~~~~~~~~~~~~~~~&\nonumber\\
&\dis=\frac{g^{2}}{4\pi}\sqrt{gH_{3}}T((4n+11.44)+i(10n+7)),&\nonumber\\
&\Pi^{r=2,3}(P_{4}=0, P_{3}=0, H_{3}, H_{8}, T, n,\sigma=+1)=~~~~~~~~~~~~~~~~~~~~~~~~~~~~~~~~~~~~~~~~~~~~~&\nonumber\\
&\dis=\frac{3g^{2}}{8\pi}c_{r=2,3}~(\frac{3}{2})^{\frac{3}{4}}
~\sqrt{\lambda_{\pm}}~\sqrt{gH_{8}}
~T((4n+11.44)+i(10n+7)),&\nonumber\\
&\Pi^{r=1}(P_{4}=0, P_{3}=0, H_{3}, T, n, \sigma=-1)=
~~~~~~~~~~~~~~~~~~~~~~~~~~~~~~~~~~~~~~~~~~~~~~~~~~~~&\nonumber\\
&\dis=\frac{g^{2}}{4\pi}\sqrt{gH_{3}}T((4n+15.62)+i(2n+9.69)),&\\
&\Pi^{r=2,3}(P_{4}=0, P_{3}=0, H_{3}, H_{8}, T, n,\sigma=-1)=~~~~~~~~~~~~~~~~~~~~~~~~~~~~~~~~~~~~~~~~~~~~~&\nonumber\\
&\dis=\frac{3g^{2}}{8\pi}c_{r=2,3}~(\frac{3}{2})^{\frac{3}{4}}
~\sqrt{\lambda_{\pm}}~\sqrt{gH_{8}}
~T((4n+15.62)+i(2n+9.69)),&\nonumber\\
&\Pi^{r=1}_{44}(P_{4}=0, P_{3}=0, H_{3}, T, n)=
~~~~~~~~~~~~~~~~~~~~~~~~~~~~~~~~~~~~~~~~~~~~~~~~~~~~~~~~~~~~~~&\nonumber\\
&\dis =\frac{ g^{2}T^{2}}{2}+
\dis\frac{g^{2}}{4\pi}\sqrt{gH_{3}}T((4n+6)+i(6n+9)),&\nonumber\\
&\Pi^{r=2,3}_{44}(P_{4}=0, P_{3}=0, H_{3}, H_{8}, T, n)=~~~~~~~~~~~~~~~~~~~~~~~~~~~~~~~~~~~~~~~~~~~~~~~~~~~~~~~~&\nonumber\\
&\dis = g^{2}T^{2}+
\dis\frac{3g^{2}}{8\pi}c_{r=2,3}~(\frac{3}{2})^{\frac{3}{4}}
~\sqrt{\lambda_{\pm}}~\sqrt{gH_{8}}
~T((4n+6)+i(6n+9)).&\nonumber
\end{eqnarray}
The sign "+" in expressions (\ref{twentyfour}) should be taken for
$r=2$ and the sign "-" corresponds accordingly to $r=3$ ($r$ is an index of the charged basis
(\ref{four})).
From Eqs. (\ref{ninteen}) and (\ref{twentyfour}) it is seen that the real parts of the PO are positive in
the ground and excited states. The imaginary parts in the expressions
$\Pi(P_{4}=0, P_{3}=0,H ,T ,n , \sigma)$ and $\Pi_{44}$ occur due to the
nonanalyticity  of a number of terms in the integrands in the r.h.s. of Eqs.
(\ref{twentytwo}),(\ref{twentythree}) for large $x \rightarrow \infty$. The
 integration contour that ensures the convergence of integrals with respect
  to $x$ results in the imaginary parts in expressions
   (\ref{twentyfour}). The imaginary part of the radiation corrections
    describes the decay of the state owing to transitions to the
 states with lower energies. The first term in the expression  $\Pi_{44}$
 is calculated by performing summation over the discrete frequencies $k_{4}$
and cannot be obtained from Eq.(\ref{twentythree}).  The second one
represents the next-to-leading term and is calculated by using a
high temperature static limit.

 The imaginary part of  $\Pi_{44}$ for $n=0$ describes the Landau damping of
 the ground state plasmon quasi-particles. It is important to note that the
 imaginary parts entering the $\Pi_{44}$ and the $\Pi(P_{4}=0, P_{3}=0,
 H , T,n=0,\sigma=+1)$ are of the
same order of magnitude. Since a spin interaction does not affect the former
correction and whereas the tachyonic state in the field is excited just due to
 the spin interaction of charged gluons, one has to conclude that the
  non-zero imaginary part of the latter function does not correspond to the
  instability of the chromomagnetic fields and also describes a usual
  damping of states at finite temperature.  Thus, to verify whether or not
  the radiation corrections in the external fields stabilize the spectrum at high
  temperature one should calculate the gluon effective mass squared
   determined by the real part of the function $\Pi(P_{4}=0, P_{3}=0,H ,
   T,n=0,\sigma=+1)$  at one-loop level. If it is positive - the spectrum,
   and hence the vacuum, is stable.

\section{Discussion}
We investigated the QCD restored phase. As it was found from the EP accounting
for the one-loop plus daisy diagrams the vacuum with non-zero Abelian
chromomagnetic fields $H_3$ and $H_8$ is favorable energetically.
It is  stable due to the magnetic masses of charged  gluons which have to be
included into consideration when the value of the background fields is
estimated. These masses are computed from the gluon one-loop polarization
operator in the external fields. As it  occurs, the charged gluon spectrum
in the fields
accounting for the tree plus the one-loop corrections
is stable in a wide
interval of temperature above the $T_d$.   Important notice, in the field
presence  the gluon magnetic mass $ m_{magn.}^2 \sim  g^2 (gH)^{1/2} T $ is
generated in one-loop order, in contrast to what happens   in the case of
the trivial vacuum where the mass $m_{magn.}^2 \sim g^4 T^2 $ is a
non-perturbative effect \cite{Kal}, \cite{Buchmuller}, \cite{Karsch}.
Since the order of the magnetic vacuum field is $ ( gH)^{1/2} \sim g^2 T$,
the order of the magnetic mass squared is $\sim g^4 T^2$. That is,
the order of the magnetic mass is the same in both calculation methods.
Clearly that the former case is also non-perturbative because the field is
taken into account exactly through the Green functions. If one accounts for
the magnetic field perturbatively, zero result follows \cite{EP}.
Of course, the stabilization of the gluon spectrum by radiation corrections
in the fields is an interesting  fact which could not be expected beforehand.
It was observed already in $SU(2)$ gluodynamics \cite{SkSt2} and here for
the $SU(3)$ gauge group. One may believe that this is the case for other
non-Abelian gauge groups and therefore the stabilization is the reflection
of the intrinsic dynamics of  fields. Note here that other mechanism of the
field stabilization was discussed in Ref.\cite{SVZ} which takes into account
the generation of the electrostatic gauge field potential
( $A_0$ condensate). But this picture was not investigated consistently
since the common generation of  the $A_0$ and magnetic  fields has not
been  considered. Some aspects of the influence of the $A_0$ condensate on
magnetic field have been investigated recently in one-loop order in
Ref. \cite{MO}.

Consider in more detail the values of the gluon magnetic
mass determined in different calculations and compare these  with the value
of the vacuum magnetic  field. In recent paper \cite{MO} in $SU(2)$
gluodynamics  to stabilize the  one-loop effective potential the gluon
magnetic mass $m_{magn.}$ of  order $\sim c g^2 T$ was introduced on
heuristic grounds. Then, in particular, it has been found that for
$ m_{magn.} \ge 0.388 g^2 T$ the effective potential has a global minimum
at $H = 0$. Hence it was concluded that a sufficiently heavy magnetic mass
leads to the trivial vacuum of the deconfinement phase. This critical value
is close to the magnetic masses determined in a number of lattice simulations
: 0.505 $g^2 T$ \cite{Cu}, 0.360 $g^2 T$ \cite{Ph}.
It is interesting to compare our result for the magnetic mass identified
with the effective mass of gluon
$M^2_{eff.}(H) = 11,44 \frac{g^2}{4\pi} (gH)^{1/2} T - gH$  and the field
$(gH)^{1/2} = \frac{g^2}{2\pi} T$ for $SU(2)$ sector. Simple estimate gives
$M_{eff.} = 0.345 g^2 T$ that is close to the value derived by Philipsen.
For this values we observed the stabilization of the magnetized vacuum.
On the other hand, this value is insufficient to have a zero vacuum field,
if the approach of the paper \cite{MO} is adopted.
Moreover, if one takes into account the structure of the magnetic
mass,
$m^{2}_{magn.}\sim \sqrt{gH}g^{2}T$, there is no possibility to
have zero for the generated field.
So, we believe that
the magnetized vacuum has to be considered seriously not an artificial
mathematical fact.

Other important point which we are going to discuss is the
influence of higher loop contributions. First we note that the one
loop plus daisy graphs account for the long distance contributions
and give the main effects. It was realized already in the related
problem on the electroweak phase transition in strong magnetic
fields \cite{SklBord}, \cite{SD}. The results obtained within this EP are in a good
agreement with that of found in the nonperturbative approach Refs.  \cite{Sh1},  \cite{Sh2}
( see
also recent survey \cite{Grasso}). The most important feature of this
approximation is that the EP is real at sufficiently high
temperatures and therefore the spontaneously generated magnetic
fields are stable. We believe that the higher loop corrections
will not change qualitatively the results obtained and the results on
the
stable magnetize vacuum survive. To check our results the resummations
on
the super daisy level have to be carried out. That is the problem for
future.

It is interesting to compare  our results with that of obtained in
lattice calculations by Cea and Cosmai \cite{Cea1}, \cite{Cea2}.
In the former  paper the creation of the colour Abelian chromomagnetic
was investigated by means of the lattice Schr\"{o}dinger functional.
It was observed that at $T=0$ the applied external chromomagnetic field
is
completely screened by the vacuum.  At finite temperature the applied
field is supported by the temperature and is increased with the growth
of
temperature. That is in correspondence with our calculations.
In the latter paper
the influence of the external fields on the deconfinement phase transition
has been investigated and an intimate connection between Abelian
chromomagnetic field and colour confinement was observed.
This interesting result is not directly related with obtained in the present
paper because we do not consider the field as external one. From our results
it follows that in the deconfinement phase the Abelian chromomagnetic
fields have to present. So, we have to answer how the spontaneous vacuum
magnetization affects the temperature of the phase transition.

\newpage
\begin{figure}\centering
\includegraphics[scale=0.4]{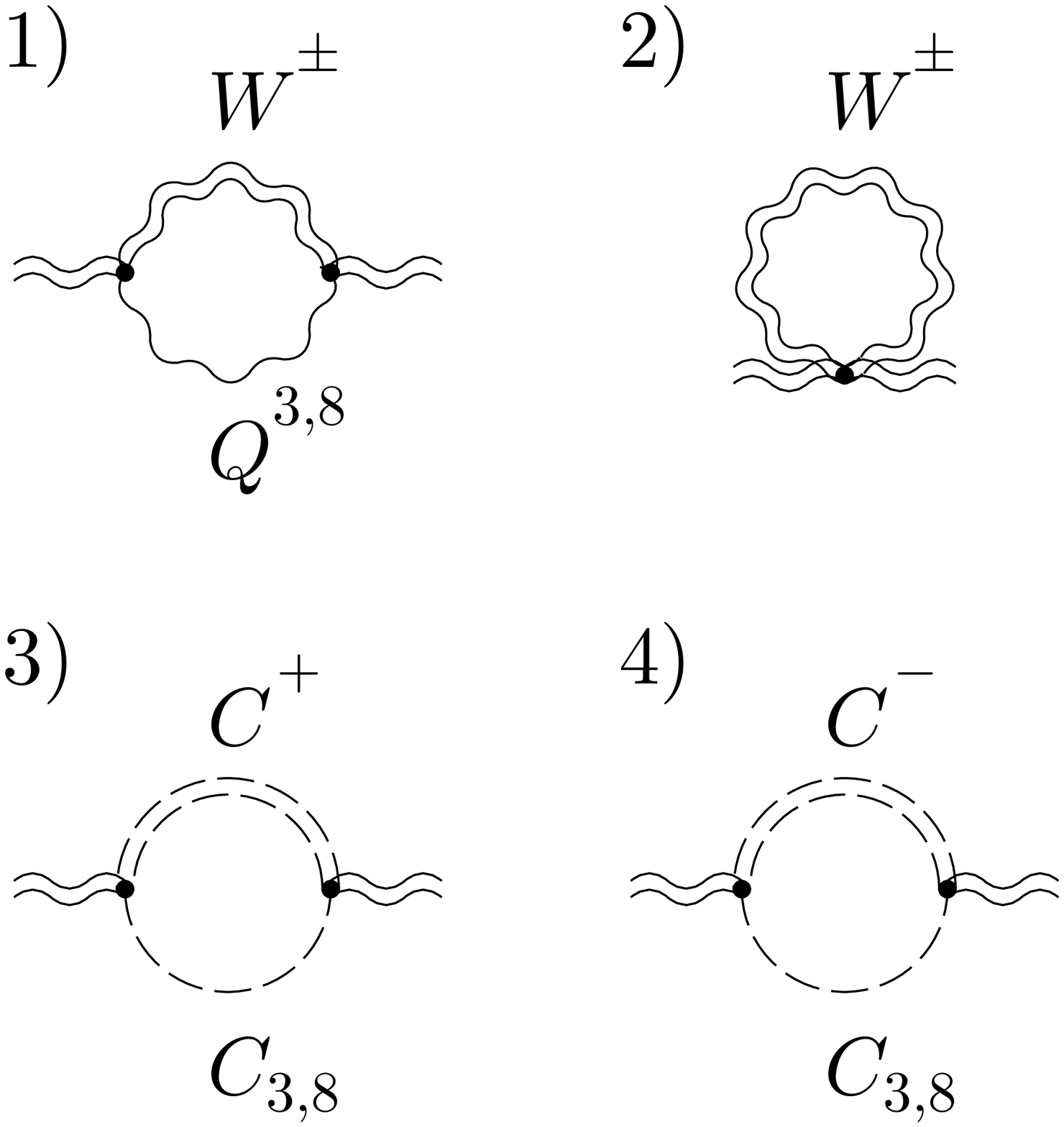}
\caption{Polarization operator of charged gluons in the one-loop approximation.} \label{Fig1}
\end{figure}

\end{document}